# Mesoscale Model for Predicting Hydrogen Damage in Face Centre Cubic Crystals

**Eugene Ogosi**[a,b], **Amir Siddiq**[b,*], **Peter Christie**[b], **Umair Bin Asim**[c], **Mehmet E Kartal**[b]

[a]*Apache North Sea, Prime Four Business Park Kingswells, Aberdeen AB15 8PU United Kingdom*
[b]*School of Engineering, University of Aberdeen, Aberdeen, AB24 3FX United Kingdom*
[c]*Department of Materials Science and Engineering, Texas A&M University, College Station, Texas 77843 USA*
[*]*Corresponding Author: amir.siddiq@abdn.ac.uk*

## ABSTRACT

A study has been performed using a crystal plasticity based finite element method to understand the effect of various stress states and crystal orientations with respect to loading direction for FCC single crystals in both hydrogenated and non-hydrogenated environment. Simulations have been performed for a variety of stress triaxilaities, Lode parmeters, crystal orientations and hydrogen concentrations. It is observed that crystal orientation has a varied effect on the influence hydrogen has on plastic deformation and void growth. Hydrogen in trap distribution at various stages of the deformation process was also found to be influenced by crystal orientation. From analyses performed, an analytical relationship between normalised void fraction and equivalent strain has been derived.

## 1. Introduction

The mechanical and corrosion resistant properties are known to be strongly dependent on the texture of the material [1]. Crystallographic orientation of steel has been found to affect hydrogen resistant properties [2][3]. Hydrogen has a deleterious effect on steel and its effect have been subject to several studies and reviews [4][5]. Specifically, experimental work has been performed on hydrogen embrittlement of austenitic stainless steel [6][7]. The hydrogen enhanced localized plasticity (HELP) theory explains how hydrogen alters dislocation interactions and influences material microstructure [5][8]. In face centred cubic (FCC) metals, with relatively low hydrogen difussivity, this has been observed to lead to local dislocation pile ups [9][10] and increased plastic flow properties [11]. Crystal orientation in relation to the rolling, normal and transverse directions have been found to influence material resistance to hydrogen induced cracking for both FCC and body centred cubic (BCC) crystal materials [12][13]. Crystal orientation parallel to the normal direction was found to improve resistance to hydrogen induced cracking [12][13]. Potirniche et al [14] performed 2D finite element simulations using a crystal plasticity based model to understand the effects of crystal orientation and loading biaxiality on void characteristics. Results showed that crystal orientation had a significant effect on void growth. Asim et al [15] observed that crystal orientation has an impact on the void growth and this effect reduces with increasing stress triaxiality. Michler and colleagues [16] found that austenite phase stability is not a guarantee that stainless steel is resistance to hydrogen embrittlement. Hua et al found that that hydrogen diffusion had a strong dependence on crystal orientation. Hydrogen mobility in 304 stainless steel crystal grains were found to be higher in (111) oriented grains when compared with (001) and (101) oriented grains[17]. Depending of crystal orientation, it was observed that hydrogen affected the plastic deformation differently [18]. The effects of crystal orientation on how hydrogen affects the properties of austenic stainless steels is investigated in this work. Simulations have been performed for a variety of stress triaxialities (0.8, 1, 1.5, 2 & 3), Lode parameters ( -1 , 0 & 1) and crystal orientations. This work builds on the findings from previous studies on hydrogen effects at different stress states [21][22] and derives an analytical relationship between strain and void growth. The material formulation is presented in section 2. Methodology is explained in section 3. Results are presented in section 4. Conclusions are presented in section 5.

## 2. Crystal plasticity model with hydrogen effects

### 2.1. Crystal plasticity theory

A crystal plasticity theory [21][22] has been extended to include hydrogen influence [19]. Deformation gradient is expressed as a product of elastic $F^e$, hydrogen $F^h$ and plastic $F^p$ components [23];

$$F = F^e F^h F^p \tag{1}$$

Marin's formulation [21] with a hydrogen component is expressed as:

$$F = V^e F^*, \quad F^* = R^e F^h F^p \tag{2}$$

$V^e$ is elastic stretch and $R^e$ is rotation. The velocity gradient $l$ in the final deformed state is given by;

$$l = \dot{F} F^{-1} \tag{3}$$

$\dot{F}$ and $F^{-1}$ are the rate of change deformation gradient and its inverse. Velocity gradient ($\tilde{L}$) in an intermediate state with elastic stretch unloaded is given by:

$$\tilde{L} = V^{e-1} l V^e = V^{e-1} \dot{V}^e + \tilde{L}^* \tag{4}$$

$V^{e-1}$ is the inverse of stretch.

$$\tilde{L}^* = \dot{R}^e R^{eT} + R^e \hat{L}^h R^{eT} + R^e F^h \bar{L}^p F^{h-1} R^{eT} \tag{5}$$

$\dot{R}^e$ is the rotation change rate and $R^{eT}$ is the transpose of $R^e$. The plastic part of the velocity gradient $\bar{L}^p$, is given by:

$$\bar{L}^p = \sum \dot{\gamma}^\alpha \, \bar{s}^\alpha \otimes \bar{m}^\alpha \tag{6}$$

$\bar{s}^\alpha$ is the direction and $\bar{m}^\alpha$ is the normal component. Combing (5) and (6):

$$\tilde{L}^* = \widetilde{\Omega}^e + \hat{L}^h + \sum_{\alpha=1}^{N} \dot{\gamma}^\alpha \, \tilde{s}^\alpha \otimes \widetilde{m}^\alpha \tag{7}$$

$\widetilde{\Omega}^e = \dot{R}^e R^{eT}$ is elastic spin. $\dot{\gamma}^\alpha$ is the rate shear strain. Sofronis provides an expression for the hydrogen deformation gradient [24]:

$$F^h = \left(1 + \frac{(c-c_o)\lambda}{3}\right) I \tag{8}$$

$c_o$ and $c$ are initial and final hydrogen concentrations. $\lambda$ is $\frac{\Delta V}{V_m}$, $\Delta V$ is volume change and $V_m$ is atomic volume. The hydrogen part of the velocity gradient $\hat{L}^h$ is:

$$\hat{L}^h = \dot{F}^h \cdot F^{h^{-1}} = \frac{1}{3}\left[\frac{3\lambda}{3+(c-c_0)}\right](c)\dot{c}I \tag{9}$$

The Second Piola-Kirchhoff stress tensor, $\tilde{S}$ is:

$$\tilde{S} = \widetilde{\mathbb{C}^e} : \widetilde{E^e} \tag{10}$$

$\widetilde{\mathbb{C}^e}$ and $\widetilde{E^e}$ are the elasticity and Green-Lagrange strain tensors. Deformation tensor splits into a symmetric, $\widetilde{D}$, and skew, $\widetilde{W}$ part:

$$\widetilde{D} = V^{e^T} dV^e = \dot{\widetilde{E}}^e + [\text{sym}(\widetilde{C}^e \widetilde{\Omega}^e) + \sum_{\alpha=1}^{N} \dot{\gamma}^\alpha \text{sym}(\widetilde{C}^e \widetilde{Z}^\alpha)] \tag{11}$$

$$\widetilde{W} = V^{e^T} wV^e = \text{skew}(V^{e^T}\dot{V}^e) + [\text{skew}(\widetilde{C}^e \widetilde{\Omega}^e) + \sum_{\alpha=1}^{N} \dot{\gamma}^\alpha \text{skew}(\widetilde{C}^e \widetilde{Z}^\alpha)] \tag{12}$$

Where $\widetilde{C}^e = R^e \widetilde{C}^e R^{eT}$ and $\widetilde{Z}^\alpha = \tilde{s}^\alpha \otimes \widetilde{m}^\alpha$.

Plastic slip evolution is given as:

$$\dot{\gamma}^\alpha = \dot{\gamma}_0^\alpha \left[\frac{|\tau^\alpha|}{\kappa_s^\alpha}\right]^{\frac{1}{m}} \text{sign}(\tau^\alpha) \tag{13}$$

$\dot{\gamma}^\alpha$ is strain rate in $\alpha$, $\dot{\gamma}_0^\alpha$ is the reference strain rate, $\kappa_s^\alpha$ is the current crystal strength of α, $\tau^\alpha$ is resolved stress and $m$ is rate sensitivity. The slip system hardens as follows;.

$$\dot{\kappa}_s^\alpha = h_0 \left(\frac{\kappa_{s,S}^\alpha - \kappa_s^\alpha}{\kappa_{s,S}^\alpha - \kappa_{s,0}^\alpha}\right) \sum_{\alpha=1}^{N} |\dot{\gamma}^\alpha|, \quad \kappa_{s,S}^\alpha = \kappa_{s,S0}^\alpha \left[\frac{\sum_\alpha |\dot{\gamma}^\alpha|}{\dot{\gamma}_{s0}}\right]^{1\backslash m'} \tag{14}$$

$\dot{\kappa}_s^\alpha$ is the current hardening rate, $h_0$ is a reference coefficient for hardening, $\kappa_{s,S}^\alpha$ is the saturation strength value and $\sum_\alpha |\dot{\gamma}^\alpha|$ is accumulated slip. $\kappa_s^\alpha(t=0)$ is the critical resolved shear stress (CRSS) of each slip system. $\kappa_{s,0}^\alpha$ $\kappa_{s,s0}^\alpha, \dot{\gamma}_{s,0}^\alpha$ and $m'$ are other plastic property defining material parameters.

## 2.2. Incorporation of hydrogen effects

Oriani proposed that hydrogen in steel will either reside in normal interstitial lattice sites (NILS) or in traps [25]. It has been discussed that diffusion of hydrogen in austenitic stainless steel is relatively slow and as such the concentration of hydrogen at material points remains unchanged during deformation, although there is a transfer of hydrogen atoms from NILS to trap sites [19]. Evidence supporting the constant hydrogen theory has been presented by Schebler [26]. Discussion on the incorporation of hydrogen influence into the model have been discussed previously [19][20] so only a summary is given here. We express the concentration of hydrogen $C_{i,bulk}$ at a given material point as:

$$C_{i,bulk} = C_L + C_{i,traps} \tag{15}$$

$C_L$ is the concentration in NILS and $C_{i,traps}$ is the hydrogen in traps before deformation. Hydrogen in traps $C_T$ during deformation increases as there is a transfer from $C_L$ to traps.

$$C_T = \theta_T \psi N_T \tag{16}$$

$\theta_T$ is hydrogen occupancy in traps, $\psi$ is the number of sites per trap and $N_T$ represent the number of traps per lattice site. $N_T$ is:

$$N_T = \frac{\sqrt{3}}{a_{fcc}} \rho \tag{17}$$

$a_{fcc}$ is lattice parameter. Evolution of dislocation density, $\dot{\rho}$ is:

$$\int_0^t \dot{\rho}\, dt = (k_1\sqrt{y}) \int_0^t /\dot{\gamma}/\, dt \tag{18}$$

$\dot{\gamma}$ is rate of change of strain. $k_1$ is a measure of immobile dislocations and $\sqrt{y}$ is the average dislocation separation length as proposed by Estrin et al [27]. Krom's formulation [28] is used to determine $C_T$:

$$C_T = \frac{1}{2}\left[\frac{N_L}{K_T} + C_{Total} + N_T - \sqrt{\left(\frac{N_L}{K_T} + C_{Total} + N_T\right)^2 - 4 N_T C_{Total}}\right] \tag{19}$$

The terms $H_i$ and $H_f$ are used to capture the hydrogen effect as previously discussed [19]

$$\kappa_{h,0}^\alpha = \kappa_{s,0}^\alpha * (1 + H_i C_{initial}) \tag{20}$$

$\kappa_{s,0}^\alpha$ is crystal strength with no hydrogen. $C_{initial}$ is hydrogen traps before plastic deformation and defined as per Caskey's relationship [29];

$$C_{initial} = f\, C_L e^{18400}/(RT) \tag{21}$$

$f$ is atoms per unit length of dislocation. 18400 J/mol is bonding energy. Crystal strength evolution in (14) is altered to include hydrogen effect as follows:

$$\dot{\kappa}_s^\alpha = h_0 \left(\frac{\kappa_{s,S}^\alpha - \kappa_s^\alpha}{\kappa_{s,S}^\alpha - \kappa_{s,0}^\alpha}\right) \sum_{\alpha=1}^N |\dot{\gamma}^\alpha| (1 + H_f C_T) \tag{22}$$

Stress triaxiality and Lode parameter are used to quantify stress states [30][31] as previously discussed by the authors [15]. Lateral displacements are controlled using a multipoint constraint (MPC) subroutine in ABAQUS finite analysis software [32] to keep stress triaxialities and Lode parameter constant at each iteration of the simulations. The relationships in sections 2.1 and 2.2 are implemented in a user material subroutine (UMAT) and ABAQUS as discussed in the next section.

## 3. Methodology

A representative volume element (RVE) is constructed using reduced integrated elements and analysed in ABAQUS finite element software [32]. Sample element with an embedded void is shown in Figure 1 with boundary conditions. The embedded void represents a nucleated defect. The relationship between strain, stress, void growth, void size, stress triaxiality and crystal orientation have been covered previously [15][33]. Only a summary is given here. Void growth is tracked by the term;

$$Normalised\ void\ fraction\ (NVF) = \frac{f}{f_0},\quad f = \frac{V_{void}}{V_{RVE}} \tag{23}$$

Initial void volume fraction, $f_0$ is given by $(4/3)\pi r^3/s^3$ where r is sphere radius, $f$ is void volume fraction, and s is side length. $V_{void}$ is the void volume and $V_{RVE}$ is the sum of solid material and void volume.

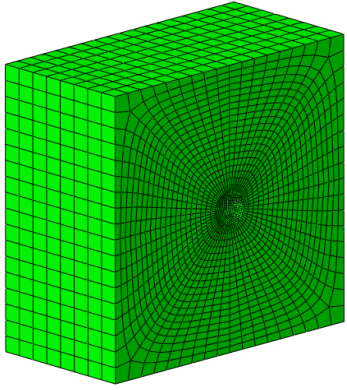

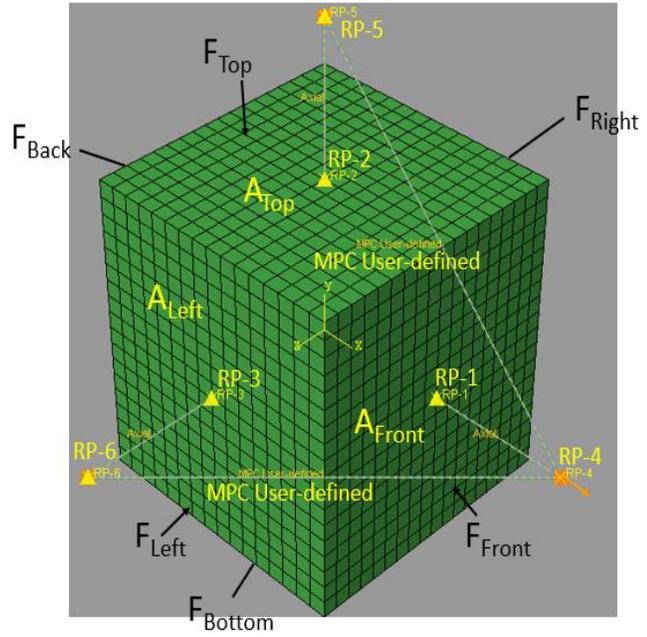

Cubical geometry - Spherical void
Void fraction ($f_o$) = 0.001
Stress triaxiality = 0.8, 1, 1.5, 2 and 3.
Lode parameter = -1, 0, 1
Hydrogen concentration = 0, 0.64%

**Figure 1: RVE sample showing boundary conditions**

Loading is done by applying a positive displacement in the x direction and the sides of the RVE are constraint to remain straight. In Figure 1, $F_{Top}$, $F_{Front}$, and $F_{Left}$ are mobile while $F_{bottom}$, $F_{back}$ and $F_{right}$ are constraint to be fixed during displacement. Lateral displacement is adjusted to keep stress triaxilities constant by applying a multipoint constraint (MPC) subroutine. Using the technique discussed by Tekoglu [34], the stress triaxiality (X) and Lode parameter (L) is held constant by satisfying;

$$\rho_{11} = \frac{\Sigma_{11}}{\Sigma_{22}}; \quad u_x = \rho_{11} \frac{A_{Front}}{A_{Top}} u_y \qquad (24)$$

$$\rho_{33} = \frac{\Sigma_{33}}{\Sigma_{22}}; \quad u_z = \rho_{33} \frac{A_{Left}}{A_{Top}} u_y \qquad (25)$$

Current RVE volume is

$$V_{RVE} = (s + u_x) * (s + u_y) * (s + u_z) \qquad (26)$$

The value of each element ($V_{e,i}$) is stored and summed to obtain

$$V_{solid} = \sum_{i=1}^{N} V_{e,i} \qquad (27)$$

Current value of void volume fraction ($f$) is then obtained by

$$f = \frac{V_{RVE} - V_{Solid}}{V_{RVE}} \qquad (28)$$

Crystal orientation is represented using Euler angles and the Kock's convention is used. If we consider a crystal (small cube) shown in Figure 2. $\psi$ represents a rotational angle in the z-axis from the original position (Z) to a first intermediate position (Z'). $\theta$ represents rotation and displacement in the y-axis from the first intermediate position (Y') to a second intermediate position (Y''). $\phi$ represents rotation and displacement in the z-axis from the second intermediate position (Z'') to the final position (Z''').

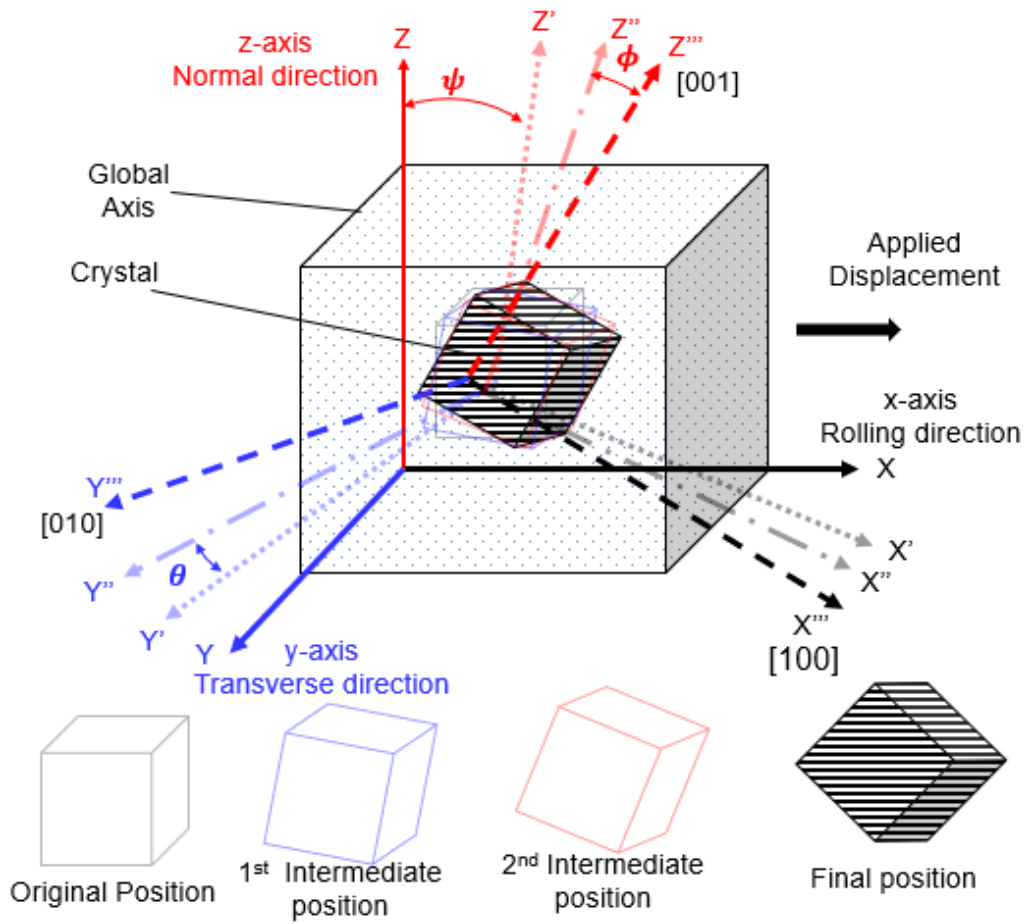

**Figure 2 Representation of crystal orientation in relation to a fixed global axis**

The crystal orientations with corresponding euler angles considered in this work are shown in Table 1.

**Table 1 Crystal orientations and Euler Angles**

| Euler Angles | $\psi$ | $\theta$ | $\phi$ |
|---|---|---|---|
| Orientation 0 (Ori 0) | 0° | 0° | 0° |
| Orientation A (Ori A) | 5° | 5° | 5° |
| Orientation B(Ori B) | 30° | 30° | 30° |
| Orientation C(Ori C) | 45° | 45° | 30° |
| Orientation D(Ori D) | 60° | 60° | 90° |
| Orientation E(Ori E) | 75° | 90° | 120° |
| Orientation F(Ori F) | 145° | 120° | 180° |

## 4. Results and discussion

### 4.1. Effect of crystal orientation and hydrogen on equivalent stress

Equivalent stress and equivalent strain curves for different crystal orientations have been derived from RVE simulations as part of this work but have not been included for brevity. It has previously been shown that the presence of hydrogen shifts the graph upwards so that the stress value required to transist from elastic to plastic deformation increases [19][35]. A similar observation was made in this work and results were consistent for a range of stress triaxialities (0.8, 1, 1.5, 2 and 3), Lode parameters (-1, 0, 1) and crystal orientations (see Table 1) considered. Crystal orientation was observed to alter the slope of the elastic modulus for both hydrogenated and non hydrogenated samples. As explained in other works [19][20], plastic work is expended primarily by dislocation activity so the presence of hydrogen also did not alter elastic modulus for various crystal orientations considered.

### 4.2. Effect of crystal orientation and hydrogen on void growth

Figure 3 shows a variation in normalised void fraction with equivalent strain and its dependence on crystal orientations. The results presented are consistent with the observations reported by Asim et al [15]. There is a shift in the graph to the left for hydrogenated samples which indicates that hydrogen promotes void growth. This has been discussed by the authors elsewhere [20]. It is noted that the effect of hydrogen varied depending on the crystal orientation. This observation was consistent for a range of stress triaxialities (0.8, 1, 1.5, 2 and 3), lode parameters (-1, 0, 1) and crystal orientations considered.

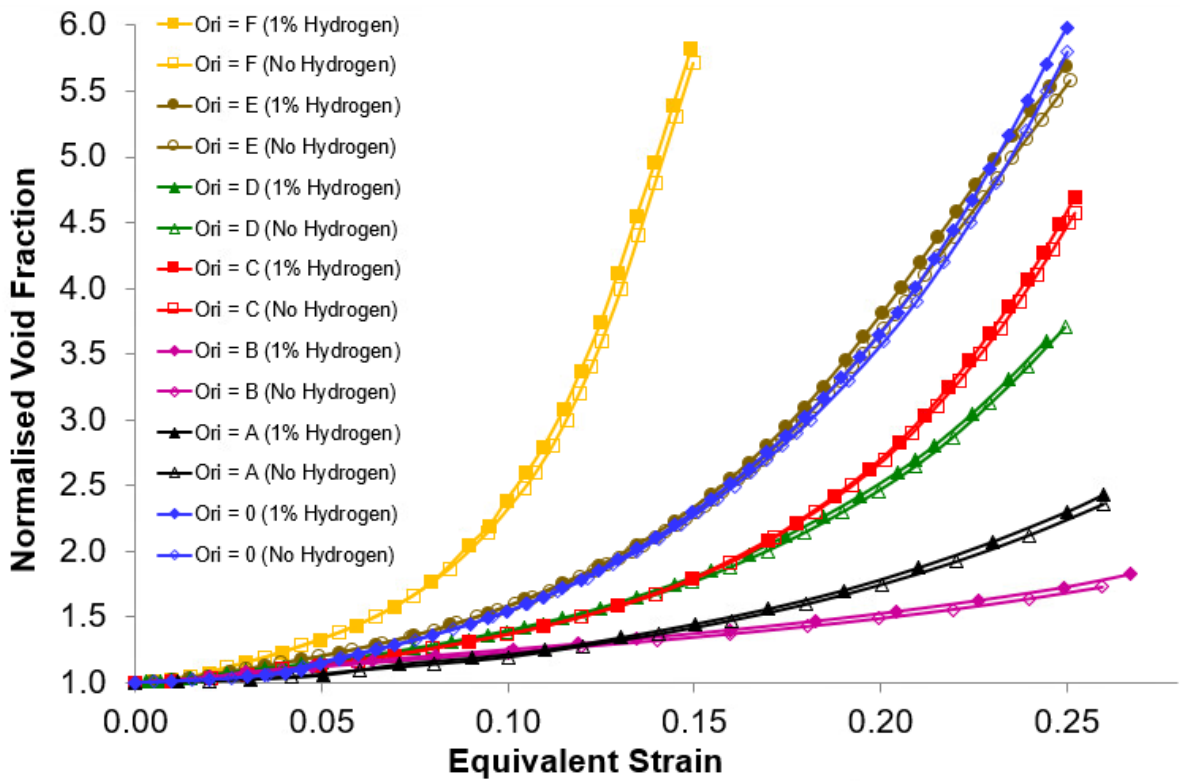

**Figure 3 Normalised Void Fraction for Different Initial Crystal Orientations, viz. stress triaxiality = 1, Lode Parameter = -1 $f_0$=0.001 and hydrogen content = 0.64%**

### 4.3. Hydrogen in traps and slip activity for different crystal orientations

Contour plots showing slip activity for the various crystal orientations are presented in Figure 4. There is evidence of a strong dependency of crystal orientation on the distribution of slip acivity which tends to concentrate around the void. The highest intensity of slip activity was observed in initial crystal orientation C and the intensity in initial crystal orientations 0, A and B were lower than C to F. The evolution of crystal orientation for different stress triaxialities and crystal orientations have been studied previously. It was found that the magnitude and direction of crystal rotation was dependent on stress triaxialities and crystal orientations [15]. Results in this work are consistent with those findings.

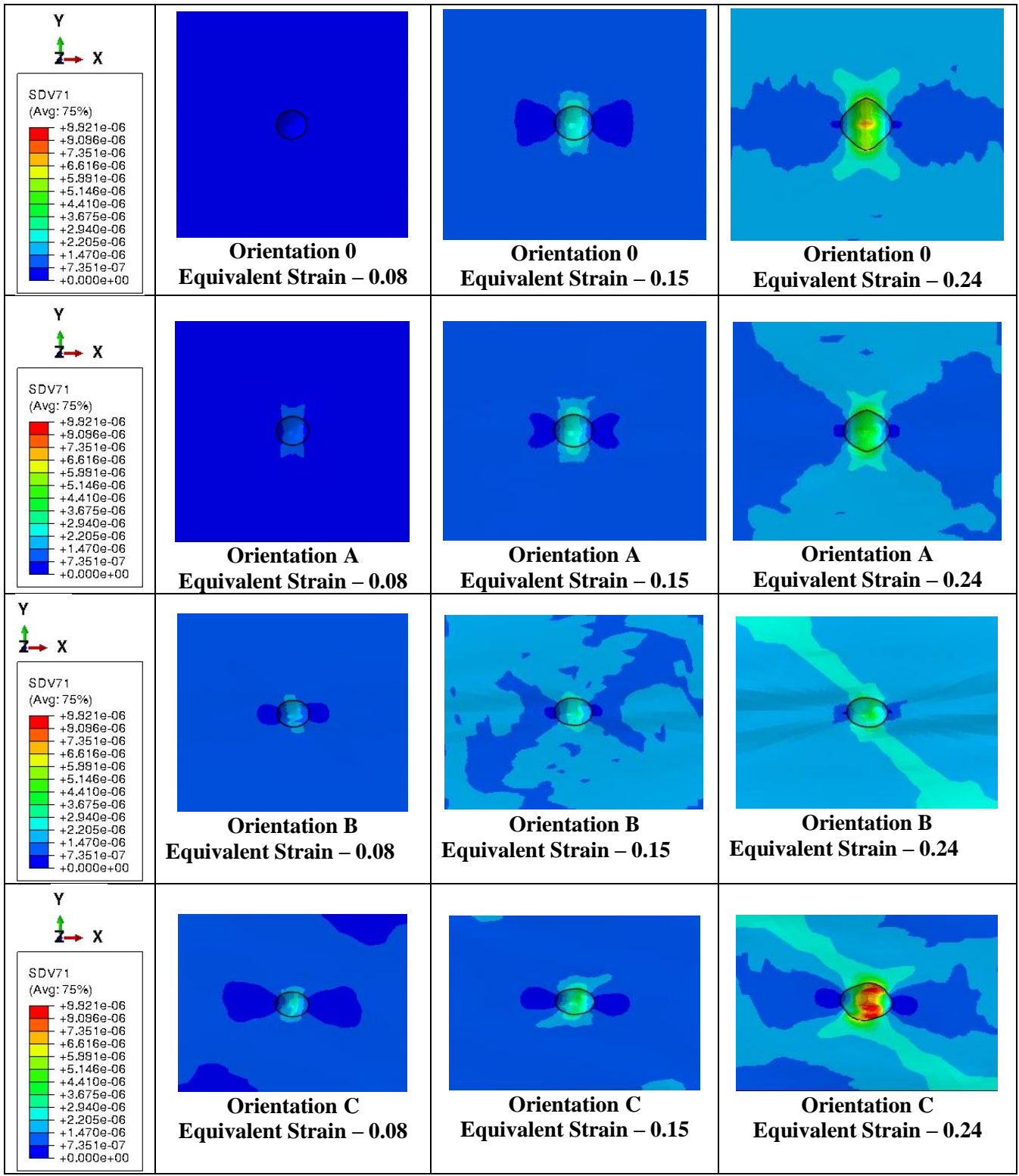

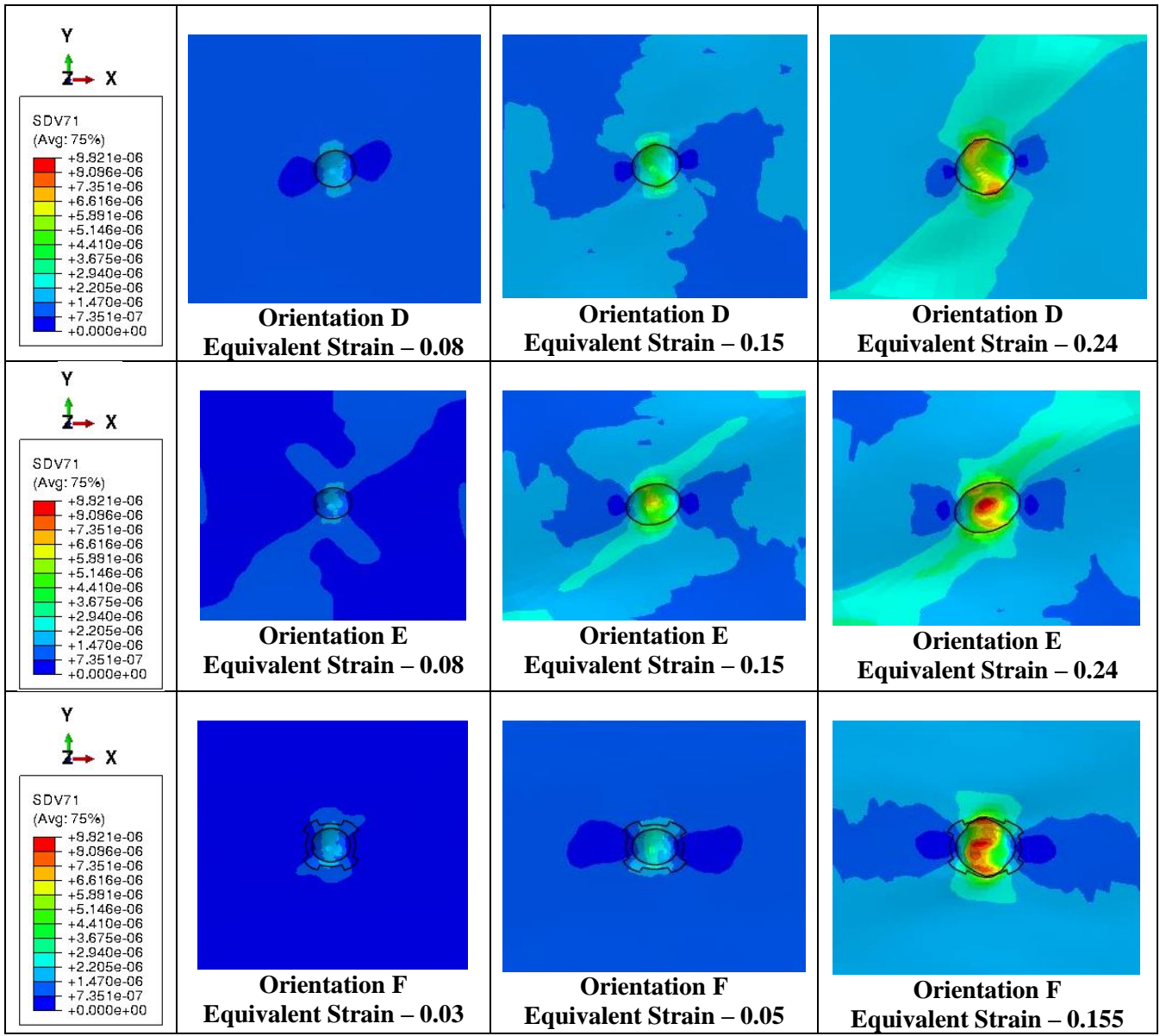

**Figure 4 Contour plots showing hydrogen in traps distribution for different crystal orientations viz. stress triaxiality = 1, Lode Parameter = -1, $f_0$=0.001 and hydrogen content is 0.64%**

4.4. Effect of crystal orientation on hydrogen in traps distribution

The distribution and magnitude of hydrogen in traps was also found to have a strong dependence on crystal orientation. Distribution followed the same pattern and intensity as slip activity. Hydrogen in traps was found to be highest in the immediate vicinity of the void as previously reported [20]. Crystal orientations 0, A and B showed a lower maximum value of hydrogen concentrated around the void relative to crystal orientations C, D, E and F. The difference in concentration of hydrogen in trap around the void may offer an explanation on the effect crystal orientation has on hydrogen influence on local plastic deformation in the area around the void and overall void growth. The relationship between hydrogen in trap concentration around the void and its influence on void growth due to the effect of crystal orientation appears to be more complex. It is worthwhile noting that the shape of the voids for the cases considered are varied and this may have an addition effect in void growth evolution if hydrogen influence is affected by void shape.

### 4.5. Proposed analytical relationship

Simulations performed have been used to derive a analytical relationship between equivalent strain and void growth evolution. An appropriate numerical expression such as linear, exponential, power law, polynomial etc was selected and used to fit the trend of experimentally simulated curves. Fitting was performed iteratively by selecting numbers that produce a close fitting to the experimentally simulated curves. This was done with the numerical computing software MATLAB. A coefficient of correlation value (or R- value) quantifies the fit of the curve and falls between 0 and 1. Fitted curves for normalised void fraction vs. equivalent strain for selected study cases returned R- values greater than 0.99 in all cases. Using the curve fitting process described, a numerical relationship below has been derived for estimating a spherical defect growth at different stress states in the presence of hydrogen

$$\text{NVF} = f(E_{eq}) = \left[aE_{eq}^{b} + 0.2\beta \sin(cE_{eq}) * \exp(-E_{eq}) + 1\right] \qquad (29)$$

NVF is normalised void fraction and $E_{eq}$ is equivalent strain. Coefficients a, b and c are material constants which are functions of stress triaxiality and Lode parameter. $\beta$ is a parameter that captures the effect of hydrogen on normalised void fraction and is directly proportional to hydrogen concentration. For a variety of stress triaxialities and Lode parameters considered, there was a close match between RVE simulations and results from the analytical model. This is evidence that the analytical model proposed can be used to predict the evolution of spherical defects for austenitic stainless-steel crystals with or without exposure to hydrogen.

## 5. Conclusions

The effect of hydrogen influence on void growth in a single crystal of austenitic stainless steel has been studied using a crystal plasticity theory and an analytical formulation has been derived. The following conclusions have been drawn from this study:

1. Crystal orientation relative to applied displacement is found to affect the influence of hydrogen on plastic deformation and void growth. This effect was found for a variety of stress triaxialities and Lode parameters considered.
2. The distribution and magnitude of hydrogen in traps, especially in the vicinity of the void, was also found to have a strong dependence on crystal orientation. The variation in hydrogen in trap distribution around the void provides evidence that there is an effect of intial crystal orientation on hydrogen influence on fracture processes.
3. The evolution of crystal orientation during plastic deformation is affected by the presence of hydrogen although there was no evidence that the general pattern of rotation is significantly affected.
4. A analytical relationship between equivalent strain and the fracture process of void growth influenced by stress triaxialities, Lode parameter and crystal orientation is presented. This relationship will be useful for engineers or material scientists that wish to estimate how defects can evolve in austenitic stainless steel crystals subject to various service loads.

## 6. Future Challenges and Opportunities

The use of mesoscale models to predict material performance provide the advantage of replicating experimental and real life conditions. These models can be useful in predicting hydrogen damage, which is a complex phenomena [36]. For metals with a FCC crystal structure, it has been explained that bulk hydrogen transport is limited due to the slow diffusive properties of hydrogen in austenite. For these materials, the interaction of hydrogen with microstructural features and the effects on the material properties is more relevant. A predictive framework based on finite element methods which incorporates the effects of stress states and crystal orientations for metals with FCC crystal structure will be of practical use to engineers. Suggestion for future work and areas of improvements are as follows:

1. Hydrogen influence is believed to be linked to activation of individual slip systems which occur differently depending on crystal orientation. This will have relevance in understanding why certain orientations favour a higher hydrogen influence on mechanical properties compared with others and should be investigated.
2. A study into the effects of hydrogen on FCC polycrystals with consideration of grain boundaries and the interaction of differently oriented crystals will be beneficial.

## Acknowledgements

This contribution is dedicated to Professor Siegfried Schmauder on the occasion of his 65th birthday. I have spent many happy hours under his supervision and in collaboration with Professor Schmauder, and it has been highly productive, educational, rewarding and enjoyable to do so. Always thankful to his continued guidance and support.

## REFERENCES

[1] S. I. Wright and D. P. Field, "Recent studies of local texture and its influence on failure," *Mater. Sci. Eng. A*, vol. 257, no. 1, pp. 165–170, 1998.

[2] V. Venegas, F. Caleyo, T. Baudin, J. H. Espina-Hernández, and J. M. Hallen, "On the role of crystallographic texture in mitigating hydrogen-induced cracking in pipeline steels," *Corros. Sci.*, vol. 53, no. 12, pp. 4204–4212, 2011.

[3] M. Masoumi, C. C. Silva, and H. F. G. de Abreu, "Effect of crystallographic orientations on the hydrogen-induced cracking resistance improvement of API 5L X70 pipeline steel under various thermomechanical processing," *Corros. Sci.*, vol. 111, pp. 121–131, 2016.

[4] S. P. Lynch, "Progress Towards the Understanding of Mechanisms of Hydrogen Embrittlement and Stress Corrosion Cracking," *NACE Corros. 2007 Conf. Expo*, no. 07493, pp. 1–55, 2007.

[5] I. M. Robertson, H. K. Birnbaum, and P. Sofronis, *Hydrogen Effects on Plasticity*, vol. 15, no. 09. Elsevier, 2009.

[6] M. Hatano, M. Fujinami, K. Arai, H. Fujii, and M. Nagumo, "Hydrogen embrittlement of austenitic stainless steels revealed by deformation microstructures and strain-induced creation of vacancies," *Acta Mater.*, vol. 67, pp. 342–353, 2014.

[7] Y. Mine and T. Kimoto, "Hydrogen uptake in austenitic stainless steels by exposure to gaseous hydrogen and its effect on tensile deformation," *Corros. Sci.*, vol. 53, no. 8, pp. 2619–2629, Aug. 2011.

[8] P. Birnbaum, H. K., Sofronis, "Hydrogen-enhanced localized plasticity—a mechanism for hydrogen-related fracture," *Mater. Sci. Eng. A*, vol. 176, no. 1–2, pp. 191–202, 1994.

[9] Y. Yagodzinskyy, T. Saukkonen, S. Kilpeläinen, F. Tuomisto, and H. Hänninen, "Effect of hydrogen on plastic strain localization in single crystals of austenitic stainless steel," *Scr. Mater.*, vol. 62, no. 3, pp. 155–158, 2010.

[10] Y. Yagodzinskyy, E. Malitckii, T. Saukkonen, and H. Hanninen, "Hydrogen-induced strain localization in austenitic stainless steels and possible origins of their hydrogen embrittlement," in *2nd International Conference on Metals and Hydrogen*, 2014, no. May, pp. 203–213.

[11] D. P. Abraham and C. J. Altstetter, "Hydrogen-enhanced localization of plasticity in an austenitic stainless steel," *Metall. Mater. Trans. A*, vol. 26, no. 11, pp. 2859–2871, 1995.

[12] G. P. Potirniche *et al.*, "Role of crystallographic texture on the improvement of hydrogen-induced crack resistance in API 5L X70 pipeline steel," *Int. J. Hydrogen Energy*, vol. 42, no. 3, pp. 14786–14793, 2017.

[13] M. Béreš *et al.*, "Role of lattice strain and texture in hydrogen embrittlement of 18Ni (300) maraging steel," *Int. J. Hydrogen Energy*, vol. 42, no. 21, pp. 14786–14793, 2017.

[14] G. P. Potirniche, J. L. Hearndon, M. F. Horstemeyer, and X. W. Ling, "Lattice orientation effects on void growth and coalescence in fcc single crystals," *International Journal of Plasticity*, vol. 22, no. 5. pp. 921–942, 2006.

[15] U. Asim, M. A. Siddiq, and M. Demiral, "Void growth in high strength aluminium alloy single crystals: A CPFEM based study," *Model. Simul. Mater. Sci. Eng.*, vol. 25, no. 3, p. 035010, 2017.

[16] T. Michler, C. San Marchi, J. Naumann, S. Weber, and M. Martin, "Hydrogen environment embrittlement of stable austenitic steels," *Int. J. Hydrogen Energy*, vol. 37, no. 21, pp. 16231–16246, 2012.

[17] Z. Hua, B. An, T. Iijima, C. Gu, and J. Zheng, "The finding of crystallographic orientation dependence of hydrogen diffusion in austenitic stainless steel by scanning Kelvin probe force microscopy," *Scr. Mater.*, vol. 131, pp. 47–50, 2017.

[18] E. G. Astafurova *et al.*, "Hydrogen-enhanced orientation dependence of stress relaxation and strain-aging in Hadfield steel single crystals," *Scr. Mater.*, vol. 136, pp. 101–105, 2017.

[19] E. I. Ogosi, U. B. Asim, M. A. Siddiq, and M. E. Kartal, "Modelling Hydrogen Induced Stress Corrosion Cracking in Austenitic Stainless Steel," *J. Mech.*, vol. 36, no. 2, pp. 213–222, 2020.

[20] E. Ogosi, A. Siddiq, U. B. Asim, and M. E. Kartal, "Crystal Plasticity based Study to Understand the Interaction of Hydrogen , Defects and Loading in Austenitic Stainless Steel Single Crystals," *Int. J. Hydrogen Energy*.

[21] E. B. Marin, "On the formulation of a crystal plasticity model.," Sandia National Laboratories, 2006.

[22] A. Siddiq and S. Schmauder, "Simulation of hardening in high purity niobium single crystals during deformation," *Steel Grips, J. steel Relat. Mater.*, vol. 3, no. 4, pp. 281–286, 2005.

[23] R. Hill and J. R. Rice, "Constitutive analysis of elastic-plastic crystals at arbitrary strain," *J. Mech. Phys. Solids*, vol. 20, no. 6, pp. 401–413, 1972.

[24] H. K. Birnbaumt and P. Sofronis, "Mechanics of the hydrogen-dislocation-impurity interactions-I. Increasing shear modulus," *J. Mech. Phys. Solids*, vol. 43, no. 1, pp. 49–90, 1995.

[25] R. A. Oriani, "Hydrogen Embrittlement of Steels," *Annu. Rev. Mater. Sci.*, vol. 8, no. 1, pp. 327–357, 1978.

[26] G. Schebler, "On the mechanics of the hydrogen interaction with single crystal plasticity," University of Illinois, 2011.

[27] Y. Estrin and H. Mecking, "A unified phenomenological description of work hardening and creep based on one-parameter models," *Acta Metall.*, vol. 32, no. 1, pp. 57–70, 1984.

[28] A. Krom, "Numerical Modelling of Hydrogen Transport of Steel," 1998.

[29] G. R. J. Caskey, "Hydrogen Solubility in Austenitic Stainless Steels," *Scr. Metall.*, vol. 34, no. 2, pp. 1187–1190, 1981.

[30] T. Luo and X. Gao, "On the prediction of ductile fracture by void coalescence and strain localization," *J. Mech. Phys. Solids*, vol. 113, pp. 82–104, 2018.

[31] C. Tekoglu, J. W. Hutchinson, and T. Pardoen, "On localization and void coalescence as a precursor to ductile fracture," *Philos. Trans. R. Soc. A Math. Phys. Eng. Sci.*, vol. 373, no. 2038, 2015.

[32] Dassault Systèmes Simulia Corp, "ABAQUS 6.18." Providence, p. 2018, 2018.

[33] U. B. Asim, M. A. Siddiq, and M. E. Kartal, "Representative volume element (RVE) based crystal plasticity study of void growth on phase boundary in titanium alloys," *Computational Materials Science*, vol. 161. pp. 346–350, 2019.

[34] C. Tekoglu, "Representative volume element calculations under constant stress triaxiality, Lode parameter, and shear ratio," *Int. J. Solids Struct.*, vol. 51, no. 25–26, pp. 4544–4553, 2014.

[35] E. I. Ogosi, U. Asim, M. A. Siddiq, and M. E. Kartal, "Hydrogen Effect on Plastic Deformation and Fracture in Austenitic Stainless Steel," *NACE Corros. Conf. Expo*, pp. 1–19, 2020.

[36] O. Barrera, D. Bombac, Y. Chen, T. D. Daff, P. Gong, and D. Haley, "Understanding and mitigating hydrogen embrittlement of steels : a review of experimental , modelling and design progress from atomistic to continuum," *J. Mater. Sci.*, vol. 53, no. 9, pp. 6251–6290, 2018.